# Antibunching and Photon Blockade in a coupled Single Quantum Dot-Cavity System.

Paolo Schwendimann and Antonio Quattropani

Institute of Physics. Ecole Polytechnique Fédérale de Lausanne.

CH 1015 Lausanne-EPFL, Switzerland

**Abstract:** In this note we present some results concerning photon blockade and antibunching in a system consisting of a quantum dot embedded in a microcavity. We give analytic conditions for resonant and non-resonant photon blockade, valid for small values of the external pump amplitude. Starting from these results, we discuss the quantum states characteristics of the system within a master equation formalism and highlight quantities like stationary second order correlations both static and dynamic and one-photon purity.

The physics of the interaction between atoms and the quantized electromagnetic field has a long history beginning with the pioneering papers by Einstein. Confining atoms in a cavity has led in the fifties, to the realization of the laser and to the emerging of Quantum Optics and Cavity Quantum Electrodynamics (CQED) [1] as autonomous research fields. In more recent years the embedding of Quantum Dots in microcavities has enlarged the field of possible applications of CQED to solid state systems. Furthermore, the interest in fundamental quantum physics and in the realization of Quantum Computers has focused on systems of one or few emitters in very good quality microcavities. One of the goals of the research in this field is to construct one-photon states. These states are expected to appear in presence of particular quantum effects like photon antibunching. Several approaches have been proposed in order to generate these states in different architectures and exploiting different material characteristics. In particular, the effect known as photon blockade [2] has been intensively studied in this context. Photon blockade is characterized by the vanishing of the second order stationary photon correlation implying that no more than one photon is present in the cavity. The vanishing of the stationary correlation and the photon blockade have been discussed theoretically and realized experimentally in a system consisting of two coupled cavities at least one of which contains a non-linear element like a Kerr crystal or one two-level atom [2-13]. The cavity-cavity coupling plays a fundamental role in these systems, the blocking being due to quantum interference between the transition paths of the photons in the two cavities.
The vanishing of the second order correlation has also been shown to occur in the case of one cavity mode interacting resonantly with one or more two-level atoms or a semiconductor quantum dot [14-21]. The presence of one photon state in one as well as in a many atom system in a cavity has been discussed in [20, 21]. Motivated by the renewed interest in photon blockade and the generation of one-photon states, in this report we discuss in some detail the onset of photon blocking in a cavity containing one or several atoms, which may or may not be in resonance with the cavity mode. From the point of view of the theory, as already stated above, photon blockade is characterized by the vanishing of the stationary second order photon correlation $g^{(2)}(0) = \langle a^+ a^+ a a \rangle / \langle a^+ a \rangle^2$. As a consequence, the one time second order photon correlation shows perfect antibunching i.e. its value at $t=0$ is zero. Furthermore, one-photon states will be produced in the cavity mode. In the experiment, the measured values of $g^{(2)}(0)$ are often found to be smaller than one but larger than zero and correspondingly a non-perfect antibunching is present. In these cases, it is agreed to assume [20, 21] that when $g^{(2)}(0) < 0.5$ it is highly probable that one-photon states are produced, while the presence blocking is not guaranteed. Indeed, as we shall show later on, such a situation may be reproduced in the theory for a specific choice of the system parameters. The presence of one-photon states for $g^{(2)}(0) < 0.5$ is discussed in terms of the one-photon purity introduced in [20]. This quantity helps identifying one-photon states in cases in which the value of the correlation is small but not zero either because of experimental limitations or because of numerical limitations. A high one-photon purity indicates that one-photon states are produced.

We start from the Hamiltonian describing the interaction of one two-level atom or of a quantum dot with a cavity mode in the presence of a pump acting on the mode

$$H = \Delta_c a^+ a + \Delta_a \sigma^+ \sigma^- + g(\sigma^+ a + \sigma^- a^+) + \alpha(a + a^+) \tag{1}$$

Here $\Delta_c = \omega_c - \omega_p$, $\Delta_{a,i} = \omega_{a,i} - \omega_p$, where $\omega_c, \omega_{a,i}, \omega_p$ are the mode atoms and pump frequencies respectively, $g$ is the mode-atom dipole coupling and $\alpha$ is the amplitude of the pump coupled to the mode. The evolution of the system is described by the master equation

$$i\hbar \frac{d\rho}{dt} = [H, \rho] + i(\Lambda_F \rho + \Lambda_A \rho) = L\rho, \tag{2}$$

where $H$ is the Hamiltonian (1), $L$ is the evolution super operator, and

$$\Lambda_F \rho = \kappa([a\rho, a^+] + [a, \rho a^+]), \tag{3a}$$

$$\Lambda_A \rho = \gamma([\sigma^- \rho, \sigma^+] + [\sigma^-, \rho \sigma^+]) \tag{3b}$$

describe the relaxation of mode and atoms respectively. Notice that often in the literature $\gamma$ is replaced by $\gamma/2$.

Optimum photon blocking appears when the second order stationary photon correlation $g^{(2)}(0)$ vanishes. In the following we also consider the second order stationary time dependent correlation

$$g^{(2)}_{stat}(0,t) = \frac{\langle a^+ a^+(t) a(t) a \rangle_{stat}}{\langle a^+ a \rangle^2_{stat}} \tag{4}$$

In order to be realistic, our considerations are done having in mind a specific system: one quantum dot characterized by the coupling and relaxation characteristics given in [17] referring to InAs quantum dots on a GaAs substrate placed inside one semiconductor or photonic crystal microcavity. In particular, we consider the atom-cavity coupling g = 76 $\mu eV$ and the half-maximum linewidth of the atom of $\gamma = 35 \mu eV$. The exciton transition is found at $\omega_a = 1.3117$ $eV$ while the detuning $\omega_c - \omega_a = \Delta$ between the cavity and the atom is tuned between $\Delta = -25 meV$ and $\Delta = 5.7 meV$ in the experiment. These values will be used in the calculations through the whole paper without further notice.

In principle, evidence of the photon blocking could be extracted from the solution of (2) in function of the coupling constant and the atom-cavity detuning. However, since we are interested in system with a small number of photons the same goal may be achieved by solving the Schrödinger equation for the system perturbatively assuming that the pump amplitude is very small and using it as the small parameter in a perturbative expansion [7]. The relaxation is phenomenolgically included into the Schrödinger equation but we are missing the contributions of quantum fluctuations. Solving the Schrödinger equation in the stationary regime, a homogeneous system of equations is found whose solvability condition allows to determine the values of the atom-photon coupling as well as of the detuning for which blocking is achieved. Once these values are found, we can discuss the details of the effect by solving the master equation (2) with the optimal choice for the coupling and the detuning. Since we assume that the pump amplitude is much smaller than one, we consider a situation in which at most two photons are present in the system.

$$i\frac{d}{dt}\langle +,1|\psi\rangle = (\Delta_c + \Delta_a - i(\gamma+\kappa))\langle +,1|\psi\rangle + g\sqrt{2}\langle -,2|\psi\rangle + \alpha\langle +,0|\psi\rangle \qquad (5a)$$

$$i\frac{d}{dt}\langle -,2|\psi\rangle = 2(\Delta_c - i\kappa)\langle -,2|\psi\rangle + g\sqrt{2}\langle +,1|\psi\rangle + \alpha\sqrt{2}\langle -,1|\psi\rangle \qquad (5b)$$

$$i\frac{d}{dt}\langle +,0|\psi\rangle = (\Delta_a - i\gamma)\langle +,0|\psi\rangle + g\langle -,1|\psi\rangle + \alpha\langle +,1|\psi\rangle \qquad (5c)$$

$$i\frac{d}{dt}\langle -,1|\psi\rangle = (\Delta_c - i\kappa)\langle -,1|\psi\rangle + g\langle +,0|\psi\rangle + \alpha\langle -,0|\psi\rangle + \alpha\sqrt{2}\langle -,2|\psi\rangle \qquad (5d)$$

$$i\frac{d}{dt}\langle -,0|\psi\rangle = \alpha\langle -,1|\psi\rangle. \qquad (5e)$$

We are interested in the stationary solution of (5), which are obtained by neglecting the time derivatives with respect to the relaxations in (5a-5d) but for (5e), which doesn't contain a relaxation term. Therefore, in what follows we replace it by its time integral. The stationary form of (5) is then

$$0 = \left(\Delta_c + \Delta_a - i(\gamma + \kappa)\right)\langle +,1|\psi\rangle + \alpha\langle +,0|\psi\rangle + g\sqrt{2}\langle -,2|\psi\rangle \qquad (6a)$$

$$0 = \left(\Delta_c - 2i\kappa\right)\langle -,2|\psi\rangle + g\sqrt{2}\langle +,1|\psi\rangle + \alpha\sqrt{2}\langle -,1|\psi\rangle \qquad (6b)$$

$$0 = \left(\Delta_a - i\gamma\right)\langle +,0|\psi\rangle + g\langle -,1|\psi\rangle + \alpha\langle +,1|\psi\rangle \qquad (6c)$$

$$0 = \left(\Delta_c - i\kappa\right)\langle -,1|\psi\rangle + g\langle +,0|\psi\rangle + \alpha\langle -,0|\psi\rangle + \alpha\sqrt{2}\langle -,2|\psi\rangle \qquad (6d)$$

$$\langle -,0|\psi\rangle = -i\alpha\int_0^\infty dt'\langle -,1|\psi\rangle(t'). \qquad (6e)$$

We next assume that $\alpha \ll 1$ and inserting (6e) into (6d) we obtain in (6d) a term of the order $\alpha^2$, which will be neglected. We now introduce the condition $g^{(2)}(0) = 0$, which in this context is equivalent to impose the condition, $\langle -,2|\psi\rangle = 0$. As a consequence, the terms containing $\langle -,2|\psi\rangle$ in (6) disappear and (6b) becomes

$$g\sqrt{2}\langle +,1|\psi\rangle + \alpha\sqrt{2}\langle -,1|\psi\rangle = 0. \qquad (6f)$$

simple algebra leads to

$$0 = \left(\Delta_c + \Delta_a - i(\gamma + \kappa)\right)\langle +,1|\psi\rangle + \alpha\langle +,0|\psi\rangle \qquad (7a)$$

$$0 = \left(\Delta_a - i\gamma\right)\langle +,0|\psi\rangle + \left(-\frac{g^2}{\alpha} + \alpha\right)\langle +,1|\psi\rangle \qquad (7b)$$

The existence of a solution different from zero of (7) is guaranteed within our approximation by the two conditions

$$g^2 - \gamma(\gamma + \kappa) + \Delta_a(\Delta_a + \Delta_c) = 0 \qquad (8a)$$

$$\Delta_a(2\gamma + \kappa) + \Delta_c\gamma = 0 \qquad (8b)$$

where a term of order $\alpha^2$ has been neglected in (8a). Notice, that (6f) may be interpreted as a destructive interference between the probability amplitudes $\langle -1|\psi\rangle$ and $\langle +1|\psi\rangle$.

At resonance, $\Delta_c = \Delta_a = 0$ and we obtain from (8a) the condition, which is the generalization of the blocking condition given by [14] in the limit $\gamma/\kappa \ll 1$. Starting from the system parameters defined above, we discuss now the possible experimental verification of (8a) in the resonant case. The dipole coupling $g$ as well as the atomic relaxation are fixed by the choice of the atom or quantum dot while the cavity mode relaxation can be varied by choosing of the cavity parameters. Once the system parameters are fixed, we solve (2) and look for a minimum of the second order stationary correlation as a function of the system parameters. As a consequence, (8a) determines the optimum blocking condition. By solving the master equation (2) with small pump but with a photon number larger than two, we verify that indeed (8a) defines the $g$ required for optimum blocking. The result is shown in Figure1 where we present the second order stationary correlation function $g^{(2)}(0)$ for a given value of the relaxations $\gamma$ and $\kappa$ while the coupling $g$ varies.

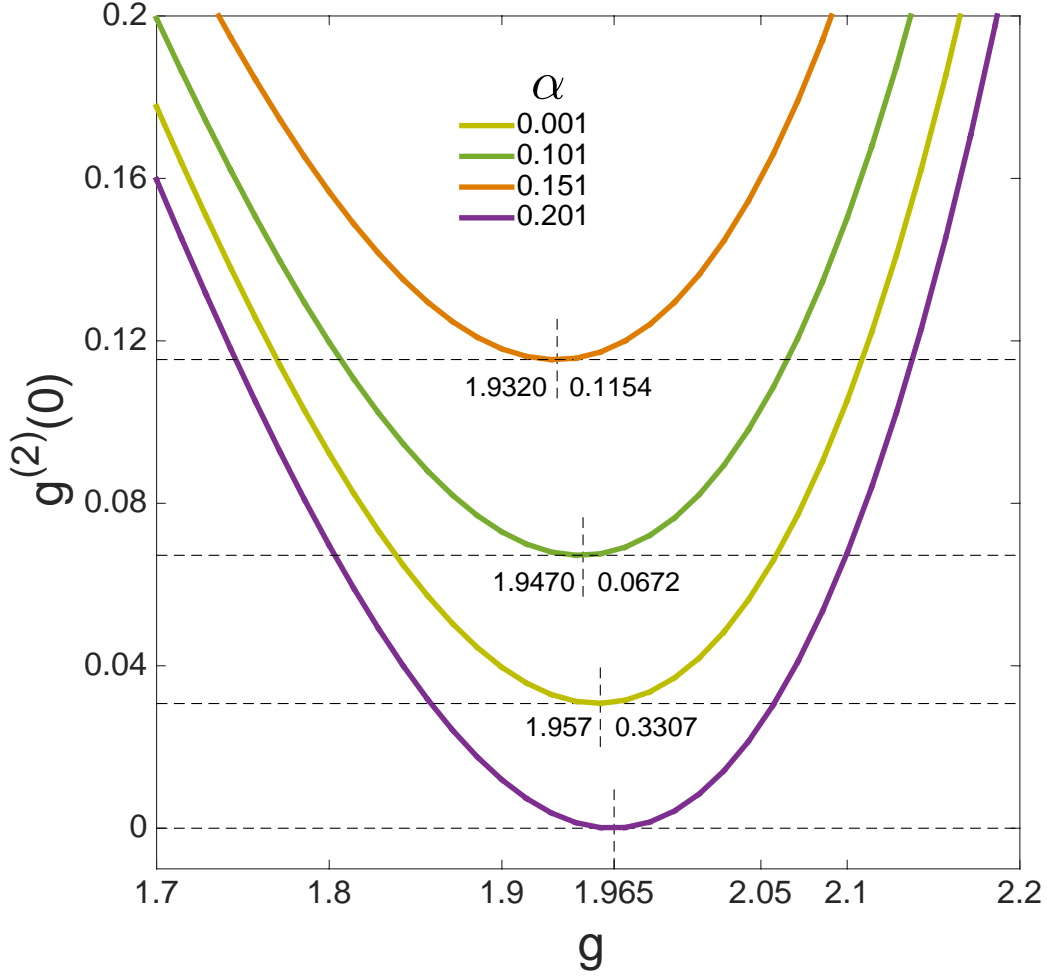

**Figure 1. Plot of the second order stationary photon correlation $g^{(2)}(0)$ at resonance ($\Delta_a = \Delta_c = 0$) as a function of the coupling $g$ for different values of the pump amplitude $\alpha$. Values and positions of the minimum of $g^{(2)}(0)$ are indicated. Energies are given in units of $\gamma$. The values of the parameters are $\kappa_{optim} = 130 \mu eV$, $g = 76 \mu eV$, , $\gamma = 35 \mu eV$ [17].**

The correlation vanishes for a value of $g$ corresponding to the blocking value from (8a). When the pump amplitude becomes larger, the minimum of the correlation becomes different from zero and eventually disappears for values of $\alpha \sim 1$. However, in an experiment, the choice of the atom or QD fixes the values of $g$ and $\gamma$ while $\kappa$ remains a free parameter. Therefore, in the resonant case, blocking is particularly effective for $\kappa/\gamma = (g/\gamma)^2 - 1$ from the solution of (8a). For different values of $\kappa/\gamma$ and with increasing $\alpha$, blocking is less effective. We present some results in Table 1,

| κ/γ | minimum g2(0) *10^5 | 10^-13*PUR |
|---|---|---|
| 3.7080 | 0.9663 | 1.4912 |
| 3.7120 | 0.3970 | 3.9208 |
| **3.7151** | **0.4343** | **6.2292** |
| 3.7200 | 0.5806 | 2.5137 |
| 3.7240 | 0.2983 | 1.0508 |
| α=0.001 | | |

| κ/γ | g2(0) minimum | 10^-5*PUR |
|---|---|---|
| 3.69 | 0.0240 | 6.2607 |
| 3.71 | 0.0236 | 6.3892 |
| 3.73 | 0.0233 | 6.4866 |
| 3.75 | 0.0232 | 6.5513 |
| **3.77** | **0.0231** | **6.5829** |
| 3.79 | 0.0232 | 6.5819 |
| 3.81 | 0.0234 | 6.5498 |
| 3.83 | 0.0237 | 6.4888 |
| 3.85 | 0.0240 | 6.4019 |
| α=0.1 | | |

**Table 1. Table of the second order stationary photon correlation $g^{(2)}(0)$ and of the one-photon stationary purity at resonance, evaluated in the framework of the master equation for two values of the pump amplitude *α* and for different values of the mode relaxation *κ* closed to the optimal value $\kappa_{optim}/\gamma = (g/\gamma)^2 - 1 = 3.7151$ ($\kappa_{optim} = 130.03 \mu eV$), solution of (8) in the resonant case $\Delta_a = \Delta_c = 0$, with $g = 76 \mu eV$ and $\gamma = 35 \mu eV$ [17].**

Besides the values of $\kappa$ and $g^{(2)}(0)$, we also report in Table 1 the values of the one-photon state purity defined as

$$PUR = \frac{p(1)}{\sum_{n>1} p(n)} \qquad (9)$$

where $p(n)$ is the probability for n photons in the mode. A large value of the purity indicates that $p(n) \ll 1$ for n>1. Notice that a high purity may be obtained also without satisfying (8). We shall discuss this point in more detail later on.

In order to discuss (8) out of resonance, we have to remember that the frequencies appearing in the Hamiltonian (1) are defined as $\Delta_{a,c} = \omega_{a,c} - \omega_p$ where $\omega_p$ is the frequency of the pump. Therefore, in the non-resonant case we have different behaviors of (8) depending on the pump frequency and on the detuning between atom and cavity. When the pump is resonant with the cavity, we have $\Delta_a \neq 0$, $\Delta_c = 0$ and (8b) cannot be satisfied. The same holds for $\Delta_a = 0, \Delta_c \neq 0$. When $\Delta_a \neq \Delta_c \neq 0$, solutions of (8) exist depending on the detuning $\Delta = \omega_c - \omega_a \neq 0$ between atom and cavity, where the value of $\omega_c - \omega_a$ is fixed once the cavity and the atom are chosen. Inserting (10) into (8b) we obtain after some simple algebra

$$\Delta_a = -\frac{\Delta}{3 + \kappa/\gamma} \qquad (10a)$$

$$\Delta_c = \Delta \frac{2 + \kappa/\gamma}{3 + \kappa/\gamma} \qquad (10b)$$

Finally, inserting (10) into (8a) and normalizing with the atomic relaxation $\gamma$, we obtain the condition for blocking

$$\frac{g^2}{\gamma^2} = (1 + \kappa/\gamma)\left(1 + \Delta^2/\gamma^2 \frac{1}{(3 + \kappa/\gamma)^2}\right) \qquad (12)$$

Notice, that (12) makes sense only when

$$\frac{g^2}{\gamma^2} \geq 1 + \kappa/\gamma \qquad (13)$$

Once the cavity and atom parameters are fixed, the only free parameter in (12) is the cavity relaxation $\kappa/\gamma$. We can determine its optimum value by solving (12) in the variable $\kappa/\gamma$. In contrast to the resonant case, the non-resonant case allows satisfying (8) or (12) by varying both the cavity relaxation and the cavity detuning. In Table 2 we present some result for the optimum blocking for various values of both $\kappa/\gamma$ and $\Delta$.

| $\Delta/\gamma$ | $\kappa/\gamma$ | minimum g2(0) *10^5 | 10^-13*PUR |
|---|---|---|---|
| 0.0 | 3.7151 | 0.6107 | 6.0978 |
| 0.2 | 3.7109 | 0.3323 | 6.8670 |
| 0.4 | 3.6984 | 0.2547 | 6.3481 |
| 0.6 | 3.6773 | 0.1309 | 6.6417 |
| 0.8 | 3.6478 | 0.4782 | 6.1148 |
| 1.0 | 3.6096 | 0.4079 | 6.1212 |
| $\alpha=0.001$ | | | |
| $\Delta/\gamma$ | $\kappa/\gamma$ | minimum g2(0) | 10^-5*PUR |
| 0.0 | 3.7151 | 0.0235 | 6.4171 |
| 0.2 | 3.7109 | 0.0235 | 6.4118 |
| 0.4 | 3.6984 | 0.0235 | 6.3960 |
| 0.6 | 3.6773 | 0.0236 | 6.3695 |
| 0.8 | 3.6478 | 0.0236 | 6.3322 |
| 1.0 | 3.6096 | 0.0237 | 6.2843 |
| $\alpha=0.1$ | | | |

**Table 2. Table of the second order stationary photon correlation $g^{(2)}(0)$ and of the one-photon purity out of resonance, evaluated in the framework of the master equation for different values of the pump amplitude $\alpha$ and for different values of the total detuning $\Delta = \Delta_c - \Delta_a$, with the corresponding values for $\kappa_{optim}$ solution of (12) and $g = 76\mu eV$ and $\gamma = 35\mu eV$ [17].**

We notice that from (10) we may also determine the optimum pump frequency for obtaining blocking. Otherwise, the same remarks as for Table 1 hold.
From the above results we follow some conditions to be satisfied in order to observe blocking. First of all we notice from Table 2 that the detuning for optimum blocking grows when the cavity relaxation becomes smaller. Since the cavity relaxation is given by the width at half height of the cavity spectrum, the detuning may be found to be outside of the cavity transmission region. Furthermore, from the pump-cavity or pump-atom detuning, we can evaluate the pump frequency, for which optimum blocking holds. Finally, equation (13) makes sure that the system is indeed out of resonance. We show in Figure 2, the expected optimum blocking out of resonance.

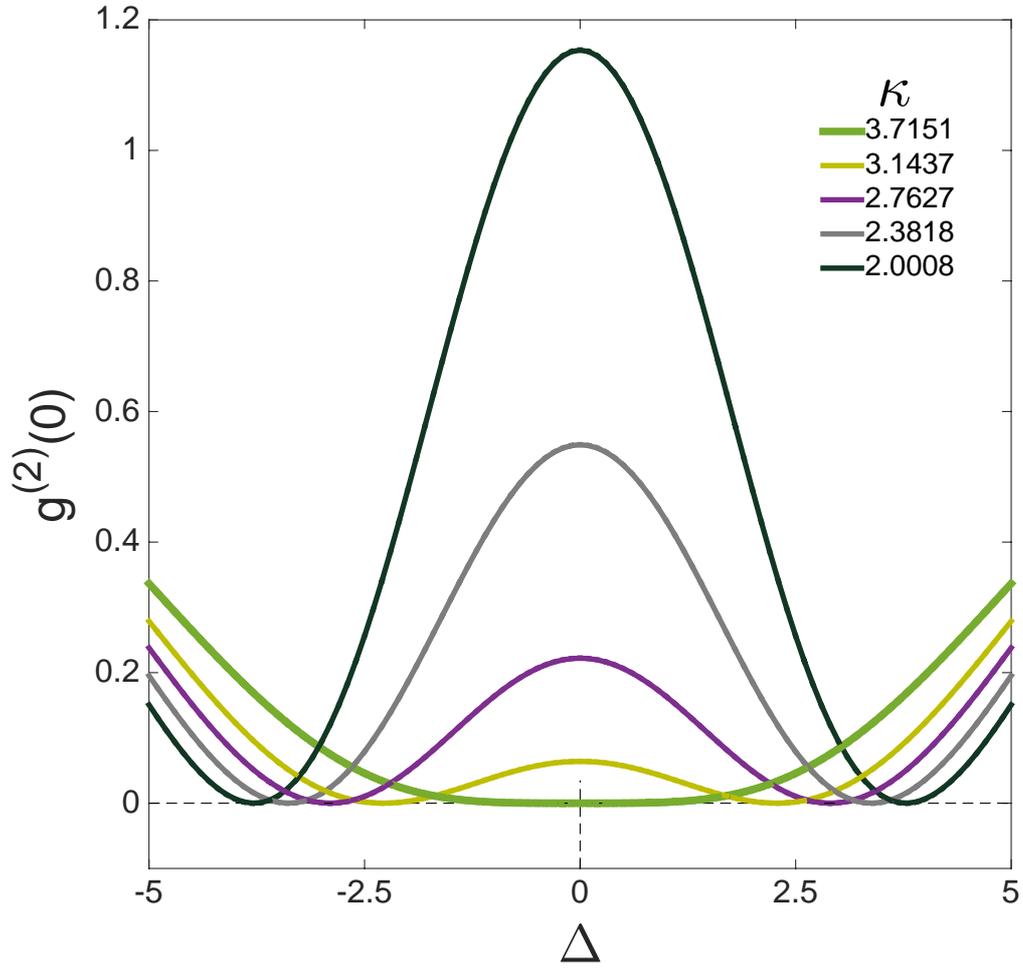

**Figure 2.** Plot of the second order stationary photon correlation $g^{(2)}(0)$ out of resonance as a function of the variable $\Delta = \omega_c - \omega_a$ for a pump amplitude $\alpha = 10^{-3}$ and for different values of the relaxation $\kappa$. Energies are given in units of $\gamma$. The values of the parameters are $g = 76\mu eV$, $\gamma = 35\mu eV$ [17].

The behavior of the second order stationary correlation in function of the pump amplitude is presented in Figure 3.

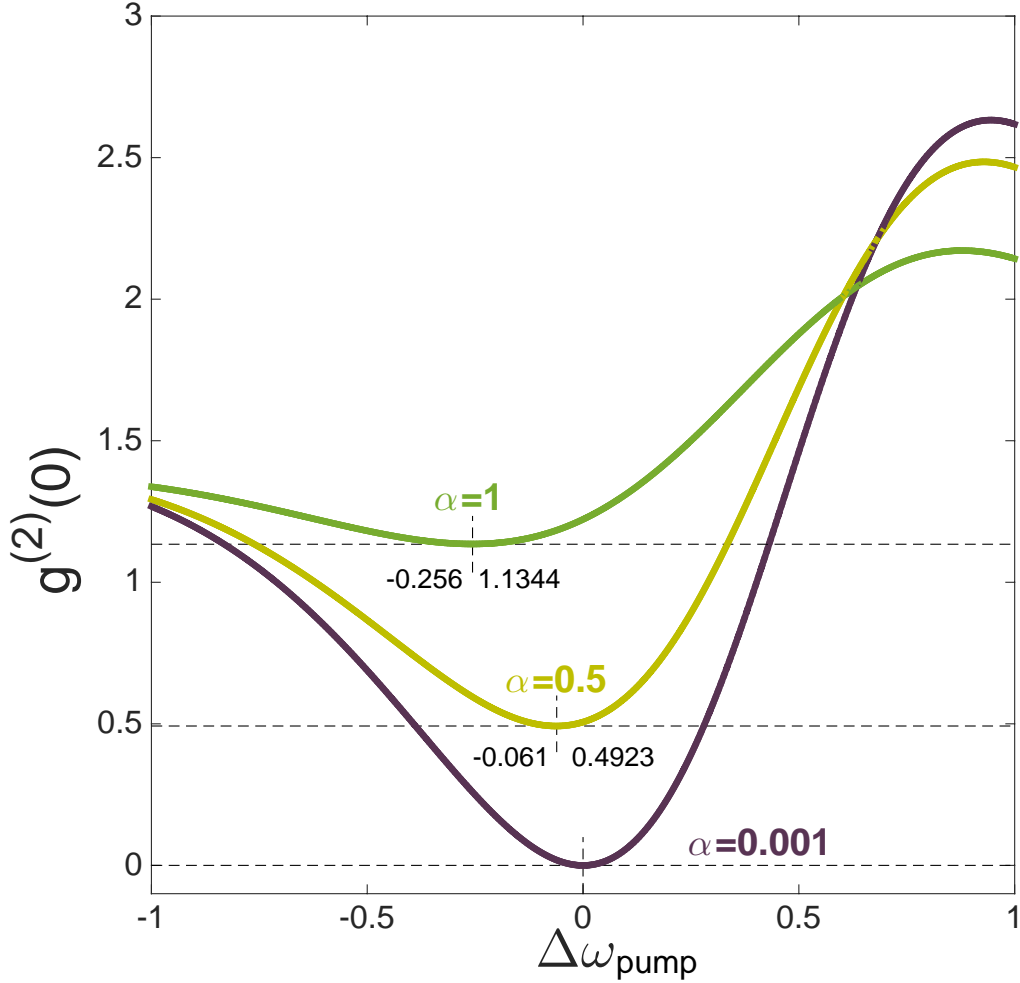

**Figure 3. Plot of the second order stationary photon correlation $g^{(2)}(0)$ out of resonance as a function of the pump frequency $\omega_{pump} = \Delta + \Delta\omega_{pump}$ for different values of the pump amplitude $\alpha$. Values and positions of the minimum of $g^{(2)}(0)$ are indicated. Energies are given in units of $\gamma$. The values of the parameters are $\kappa = 100\mu eV$, $g = 76\mu eV$, $\gamma = 35\mu eV$ [17]**.

$$\Delta_a = \omega_a - \omega_{pump} + \Delta\omega_{pump} = 2.2908 + \Delta\omega_{pump},$$
$$\Delta_c = \omega_c - \omega_{pump} + \Delta\omega_{pump} = -0.4716 + \Delta\omega_{pump}.$$

The stationary correlation shows a minimum whose position changes with growing pump amplitude. As shown in Table 2 the minimum found for small pump amplitudes is related to the high probability of having a one-photon state in the system. When the pump amplitude becomes larger, the probability of having contribution of many-photon states becomes larger. As a consequence, blocking is less effective. For large pump amplitudes, blocking is no more present. We also notice, that for small pump values, the emitted radiation shows sub-poissonian statistics. Finally in Figure 4 we present the one-time second order correlation $g^{(2)}(0,t)$ for different values of the pump amplitude

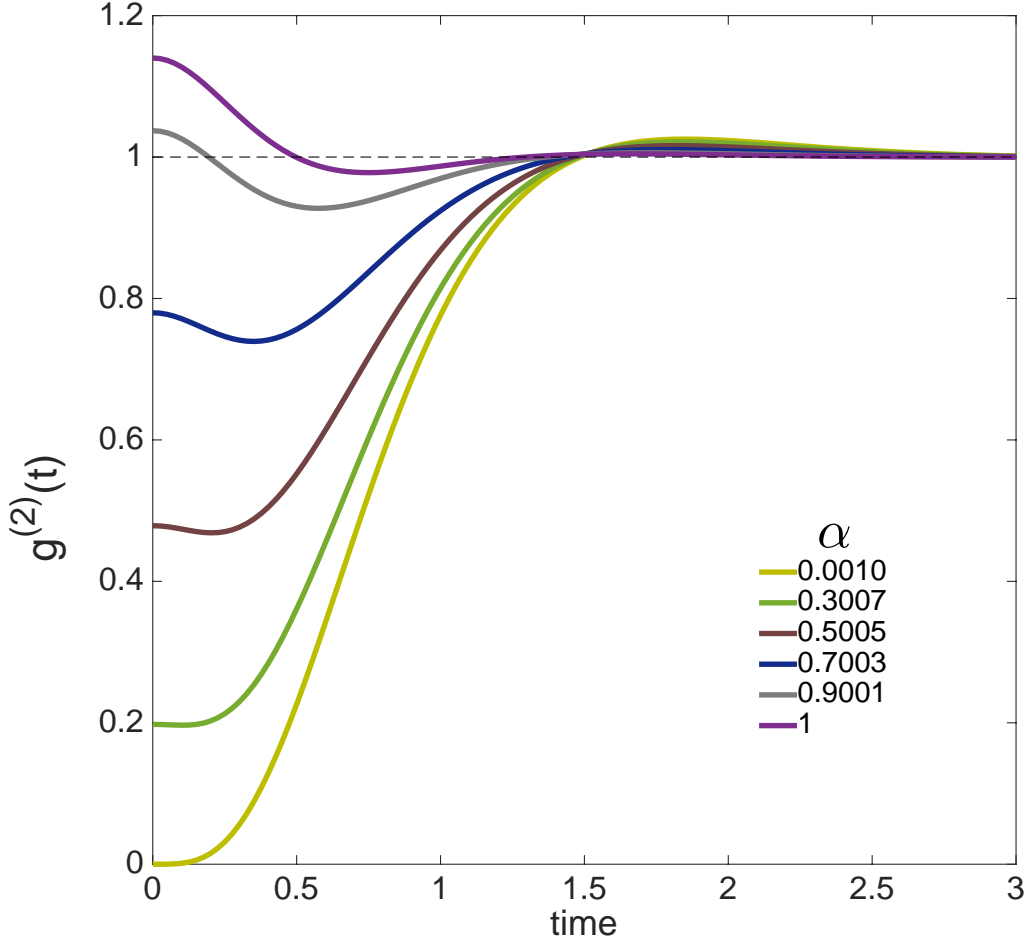

**Figure 4.** Plot of the second order stationary photon correlation $g^{(2)}_{stat}(0,t)$ at resonance as a function of time for different values of the pump amplitude $\alpha$. Energies are given in units of $\gamma$. The values of the parameters are $g = 76 \mu eV$, $\kappa_{optim} = g^2 - 1 = 3.7151$, $\gamma = 35 \mu eV$ [17].

In general, antibunching occurs when $g^{(2)}_{stat}(0,t) > g^{(2)}_{stat}(0,0)$ and $g^{(2)}_{stat}(0,0) \equiv g^{(2)}(0) < 1$, while on a sufficiently long time scale $g^{(2)}_{stat}(0,t) \to 1$. Thus, a situation for which $g^{(2)}_{stat}(0,t) < 1$ will always exhibit antibunching on some time scale [22]. As expected, for small pump values in the correlation $g^{(2)}(0,t)$ antibunching is found whose relevance diminishes with growing pump values. This behavior reproduces the one presented in Table 1 for the blocking for different values of the pump amplitude. According to our calculations, we find that blocking and antibunching are connected through the condition $g^{(2)}_{stat}(0,0) = 0$ for small values of the pump amplitude. Perfect antibunching is a consequence of the inability of one atom to emit two photons at the same time or equivalently of the fact that an atom after emission of a photon has to absorb a second photon before emitting. Since this condition implies that in our case only a single photon is present in the cavity, we conclude that a one-photon state is achieved. We call this situation perfect blocking. For larger values of the pump and the same values of the parameters, perfect blocking is no more achieved, but a minimum in $g^{(2)}(0)$ as well as a non-negligeable amount of antibunching are found. An analogous situation is found in the resonant case when the condition (8) no more holds but the pump amplitude is small. The results are presented in Figure. 5

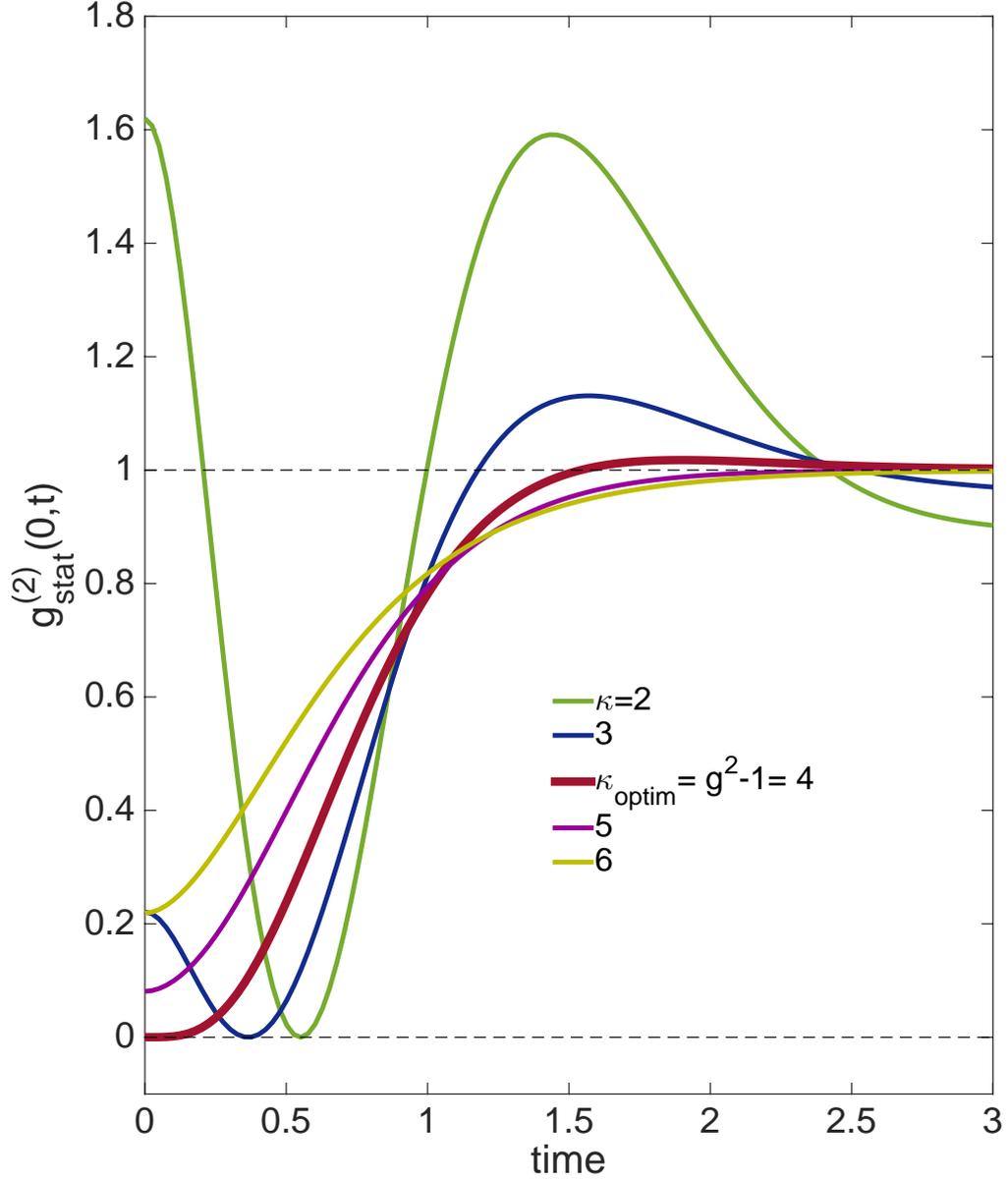

**Figure 5. Plot of the second order stationary photon correlation $g^{(2)}_{stat}(0,t)$ at resonance as a function of time for different values of the cavity relaxation $\kappa$ around the optimal value $\kappa_{optim} = g^2 - 1$. Energies are given in units of $\gamma$. The values of the parameters[17] are $g = 76 \mu eV$, $\kappa_{optim} = g^2 - 1 = 3.7151$, $\gamma = 35 \mu eV$, $\alpha = 10^{-4}$.**

As shown in Figure 5, $g^{(2)}_{stat}(0,t)$ has the following behavior for a fixed small pump amplitude as a function of the cavity relaxation: when $\kappa < \kappa_{optim} = g^2 - 1$, $g^{(2)}_{stat}(0,0) > 0$ shows oscillation, and $g^{(2)}_{stat}(0,t)$ decreases in time until for a specific time $t_0$, $g^{(2)}_{stat}(0,t_0) \approx 0$, and it converges to one for larger times. In this case, $g^{(2)}_{stat}(0,0)$ has a sub-poissonian character. When $\kappa > \kappa_{optim} = g^2 - 1$, $g^{(2)}_{stat}(0,0) < 1$, antibunching is still present but the condition (8) for blocking is no more satisfied, and oscillations disappear. We want to stress the fact that antibunching is a signature for the presence of specific quantum effects like the realization of one-photon states, which are missing when $g^{(2)}(0) \geq 1$. Concerning the presence of one-photon states we can summarize our results as follows: For very small pump amplitudes $\alpha \ll 1$, one-photon states are unambiguously found for perfect blocking. In other situations, like in the case of blocking with a larger pump intensity $\alpha < 1$ or when (8) is not satisfied, quantum effects are present in the system but the contribution of two- or three-photon states may not be negligible. We can get more information on the photon states present in the system by calculating from (2) the photon probabilities $p(n)$ for different values both of the pump and of the cavity relaxation as discussed above and relating them to the corresponding values of the stationary correlation. We find that when $g^{(2)}(0) < 0.5$ the probability of having one photon in the system $p(1)$ is some orders of magnitude larger than the probabilities $p(n)$, $n > 1$ in all situations considered above. This result is in agreement with some statements found in the literature [20, 21] concerning the relevance of one-photon states in absence of perfect blocking. A further quantity, which contains information about the relevance of one-photon states, is the one-photon purity (9) [20]. However, a criterion is missing, determining a lower limit for the significant values of the one-photon purity in a specific situation. Indeed, the values found for the one-photon purity may strongly vary (see Table1). As stated above, the value of the stationary correlation sets a limit for the observation of quantum effects because for $g^{(2)}(0) = 1$ no antibunching is found. In the case of the purity this limit is missing. It may happen that a non-negligible one-photon purity appears when $g^{(2)}(0) \geq 1$. Therefore, care must be taken in using this quantity for identifying the overwhelming presence of one-photon states.

Summarizing, the first point presented in this note is the description of non-resonant photon blockade in terms of analytical relations derived in the small pump regime. We have further discussed the quantum effects like antibunching and the emergence of one-photon states when the pump amplitude values grow or perfect blocking is not achieved.